\documentstyle[psfig]{mn}

\def\etal{{\it et al.\ }}
\def\eg{{\it e.g.\ }}

\def\spose#1{\hbox to 0pt{#1\hss}}
\def\approxlt{\mathrel{\spose{\lower 3pt\hbox{$\sim$}}
	\raise 2.0pt\hbox{$<$}}}
\def\approxgt{\mathrel{\spose{\lower 3pt\hbox{$\sim$}}
	\raise 2.0pt\hbox{$>$}}}
\def\approxpropto{\mathrel{\spose{\lower 3pt\hbox{$\sim$}}
	\raise 2.0pt\hbox{$\propto$}}}
\mathchardef\twiddle="2218

\def\multleft#1{\hbox to size{\vbox {\halign {\lft{##}\cr #1}}\hfill}\par}
\def\multright#1{\hbox to size{\vbox {\halign {\rt{##}\cr #1}}\hfill}\par}

\def\today{\ifcase\month\or January\or February\or March\or April\or May\or
      June\or July\or August\or September\or October\or November\or December\fi
      \space\number\day, \number\year}
\def\<{\thinspace}
\def\km{{\rm\thinspace km}}

\def\Mpc{{\rm\thinspace Mpc}}

\def\s{{\rm\thinspace s}}

\def\kmps{\hbox{$\km\s^{-1}\,$}}

\def\kmpspMpc{\hbox{$\kmps\Mpc^{-1}$}}

\title[Cosmological constraints from the X-ray gas mass fraction in 
relaxed clusters]
{Cosmological constraints 
from the X-ray gas mass fraction in 
relaxed lensing clusters observed with Chandra}

\author[S.W. Allen et al.]
{\parbox[]{6.in} {S.W. Allen, R.W. Schmidt and A.C. Fabian \\
\footnotesize
Institute of Astronomy, Madingley Road, Cambridge CB3 0HA
 }}
\begin{document}
\maketitle
\begin{abstract}
We present precise measurements of the X-ray gas mass fraction for a sample 
of luminous, relatively relaxed clusters of galaxies observed with the 
Chandra Observatory, for which independent confirmation of the mass results 
is available from gravitational lensing studies. Parameterizing the total 
(luminous plus dark matter) mass profiles using the model of Navarro, 
Frenk \& White (1997), we show that the X-ray gas mass fractions in the 
clusters asymptote towards an approximately constant value at a radius 
$r_{2500}$, where the mean interior density is $2500$ times the critical 
density of the Universe at the redshifts of the clusters. Combining the 
Chandra results on the X-ray gas mass fraction and its apparent redshift 
dependence with recent measurements of the mean baryonic matter density 
in the Universe and the Hubble Constant determined from the Hubble Key 
Project, we obtain a tight constraint on the mean total matter density of 
the Universe, $\Omega_{\rm m} = 0.30^{+0.04}_{-0.03}$, and measure a
positive cosmological constant, $\Omega_{\Lambda} = 0.95^{+0.48}_{-0.72}$.
Our results are in good agreement with recent, independent findings
based on analyses of anisotropies in the cosmic microwave
background radiation, the properties of distant supernovae, and 
the large-scale distribution of galaxies.
\end{abstract}


\section{Introduction}

The matter content of rich clusters of galaxies is thought to provide
a fair sample of the matter content  of the Universe  as a whole
(White \etal 1993). The observed ratio of the baryonic to  total mass
in clusters should therefore closely match the ratio of the
cosmological parameters  $\Omega_{\rm b}/\Omega_{\rm m}$, where
$\Omega_{\rm b}$ and  $\Omega_{\rm m}$ are the mean baryon and total
mass densities of the Universe, in units of the critical density.  The
combination of robust measurements of the baryonic mass fraction in
clusters with accurate determinations of $\Omega_{\rm b}$ from cosmic
nucleosynthesis calculations (constrained by the observed abundances
of light  elements at high redshifts) can therefore be used to
determine $\Omega_{\rm m}$.

This method for measuring $\Omega_{\rm m}$, which is particularly
simple in terms of its underlying assumptions, was  first highlighted
by White \& Frenk (1991) and subsequently employed by a number of
groups (\eg Fabian 1991, White \etal 1993, David, Jones \& Forman
1995; White \& Fabian 1995; Evrard 1997;  Fukugita, Hogan \& Peebles
1998; Ettori \& Fabian 1999; Bahcall \etal 2000). In general,  these
studies have found $\Omega_{\rm m}< 1$ at high significance,  with
preferred values lying in the range $\Omega_{\rm m} \sim (0.1-0.3)h^{-0.5}$.

Sasaki (1996) and Pen (1997) described how measurements of the mean
baryonic mass fraction in clusters as a function of redshift can,
in principle, be used to place more detailed constraints on cosmological
parameters, since the observed baryonic mass fraction values are
sensitive to the angular diameter distances to the clusters 
assumed in the analyses. Until now, however, systematic uncertainties 
in the observed quantities have seriously complicated the application of 
such methods.

The baryonic mass content of rich clusters of galaxies is dominated by
the X-ray emitting intracluster gas, the mass of which exceeds the mass
of the optically luminous material by a factor $\sim 6$ (\eg White
\etal  1993; David \etal 1995; Fukugita,  Hogan \& Peebles
1998). Since the X-ray emissivity of the X-ray gas is proportional to
the square of its density, the gas mass profile can be precisely
determined from the X-ray data.  With the advent of accurate
measurements of $\Omega_{\rm b}$ (\eg O'Meara  \etal 2001 and
references therein) and a precise determination of the Hubble Constant
(Freedman \etal 2001), the dominant uncertainty in determining
$\Omega_{\rm m}$ from the baryonic mass fraction in clusters 
has lain in the measurements of the total (luminous plus dark)
matter distributions in the individual clusters.

In this letter we report precise measurements of the X-ray gas mass
fraction for a sample of luminous, relatively relaxed clusters
spanning the  redshift range $0.1<z<0.5$, for which  precise,
consistent mass models have recently been determined from Chandra
X-ray data and independent gravitational lensing constraints (Allen
\etal 2001a, 2002;  Schmidt, Allen \& Fabian 2001). The agreement
between the mass measurements determined from the two independent
methods firmly limits the systematic  uncertainties in the baryonic
mass fraction measurements to $\approxlt 10$ per cent,  an accuracy
comparable to the current $\Omega_{\rm b}$ results.  With the reduced
systematic uncertainties, we show that a method  similar to those
proposed by Sasaki (1996) and Pen (1997) can be successfully applied
to the data, resulting in a tight constraint on  $\Omega_{\rm m}$ and
an interesting constraint on $\Omega_{\rm \Lambda}$. We show that the
results obtained are in good agreement with those from recent studies
of anisotropies in the cosmic microwave background radiation, the
large-scale distribution of galaxies, and the properties of distant
supernovae (\eg Jaffe \etal 2001;  Efstathiou \etal 2001).

Results on the X-ray gas mass fractions are quoted  for two default
cosmologies: SCDM with  $h = H_0/100$\kmpspMpc $= 0.5$,  $\Omega_{\rm
m} = 1$  and $\Omega_\Lambda = 0$, and $\Lambda$CDM with $h=0.7$,
$\Omega_{\rm m} = 0.3$ and $\Omega_\Lambda = 0.7$.

\section{Observations and data analysis}

\begin{table}
\begin{center}
\caption{Summary of the Chandra observations.
}\label{table:targets}
\vskip -0.2truein
\begin{tabular}{ c c c c c c c }
&&&&  \\
              & ~ &  z  & Date   &  Exposure (ks) \\
\hline
PKS0745-191      & ~ &  0.103 & 2001 Jun 16 & 17.9 \\
Abell 2390       & ~ &  0.230 & 1999 Nov 07  & 9.1  \\
%Abell 2667       & ~ &  0.233 & 2001 Jun 19 & 9.6  \\
Abell 1835       & ~ &  0.252 & 1999 Dec 12 & 19.6 \\
MS2137-2353      & ~ &  0.313 & 1999 Nov 18 & 20.6 \\
RXJ1347-1145(1)  & ~ &  0.451 & 2000 Mar 05 & 8.9 \\
RXJ1347-1145(2)  & ~ &  0.451 & 2000 Apr 29 & 10.0 \\
3C295            & ~ &  0.461 & 1999 Aug 30 & 17.0 \\
&&&& \\		    
\hline			    
\end{tabular}
\end{center}
% \parbox {7in}
%{}
%\vskip 0.5truein
\end{table}

The Chandra observations were carried out using the back-illuminated S3 
detector on the Advanced CCD Imaging Spectrometer (ACIS) between 1999 
August 30 and 2001 June 16. For our analysis we have used the 
the level-2 event lists provided by the standard Chandra pipeline 
processing. These lists were cleaned for periods of background flaring 
using the CIAO software package resulting in the net exposure times 
summarized in Table \ref{table:targets}. 

The Chandra data have been analysed using the methods described by
Allen \etal (2001a, 2002) and Schmidt \etal (2001). In brief,
concentric annular spectra were extracted from the cleaned event
lists,  centred on the peaks of the X-ray emission from the
clusters.\footnote{For  RXJ1347-1145, the data from the southeast
quadrant of the cluster were  excluded due to ongoing merger activity
in that region; Allen \etal (2002).}  The spectra were analysed using
XSPEC (version 11.0: Arnaud 1996), the  MEKAL plasma emission code
(Kaastra \& Mewe 1993; incorporating the Fe-L  calculations of
Liedhal, Osterheld \& Goldstein 1995), and the photoelectric
absorption  models of Balucinska-Church \& McCammon (1992). Only data
in the  $0.5-7.0$ keV energy range were used. The spectra for all
annuli were  modelled simultaneously, in order to determine the
deprojected X-ray gas  temperature profiles under the assumption of
spherical symmetry.

For the mass modelling, azimuthally-averaged surface brightness
profiles were constructed from background subtracted, flat-fielded
images with a $0.984\times0.984$ arcsec$^2$ pixel scale ($2\times2$
raw detector  pixels). When combined with the deprojected spectral
temperature profiles,  the surface brightness profiles can be used to
determine the X-ray gas mass profiles (to high precision) and total
mass profiles in the clusters.\footnote{The observed surface
brightness profile and a particular parameterized mass model are
together used to predict the temperature profile of the X-ray gas. (We
use the median temperature profile determined from 100 Monte-Carlo
simulations. The outermost pressure is fixed using an iterative
technique which ensures a smooth pressure gradient in these regions.)
The predicted temperature profile is rebinned to the same binning as
the spectral results and the $\chi^2$  difference between the
observed and predicted, deprojected temperature  profiles is
calculated. The parameters for the mass model are  stepped through a
regular grid of values in the $r_{\rm s}$-$\sigma$ plane (see text) to
determine the best-fit values and 68 per cent confidence limits. (The
best-fit models generally provide good descriptions of the data). The
gas mass profile is determined to high precision at each grid point
directly from the observed surface  brightness profile and model
temperature profile. Spherical symmetry and  hydrostatic equilibrium
are assumed throughout.} For this analysis, we have used an enhanced
version of the image deprojection code described by White, Jones \&
Forman (1997) with distances calculated using the code of Kayser,
Helbig \& Schramm (1997).

We have parameterized the cluster  mass (luminous plus dark matter)
profiles using a  Navarro, Frenk \& White (1997; hereafter NFW) model
with

\begin{equation}
\rho(r) = {{\rho_{\rm c}(z) \delta_{\rm c}} \over {  ({r/r_{\rm s}}) 
\left(1+{r/r_{\rm s}} \right)^2}},
\end{equation}

%$\rho(r) = \rho_{\rm c}(z) \delta_{\rm c} / [({r/r_{\rm s}})
%\left(1+{r/r_{\rm s}} \right)^2]$, 

\noindent where $\rho(r)$ is the mass density, 
$\rho_{\rm c}(z) = 3H(z)^2/ 8 \pi
G$ is the critical density for closure at redshift $z$, $r_{\rm s}$ is
the scale  radius, $c$ is the concentration parameter (with
$c=r_{200}/r_{\rm s}$) and  $\delta_{\rm c} = {200 c^3 / 3 \left[
{{\rm ln}(1+c)-{c/(1+c)}}\right]}$.   The normalizations of the mass
profiles may also be expressed in terms of  an equivalent velocity
dispersion, $\sigma = \sqrt{50} r_{\rm s} c H(z)$  (with $r_{\rm s}$
in units of Mpc and $H(z)$ in \kmpspMpc).

In determining the results on the X-ray gas mass fraction,  $f_{\rm
gas}$, we have adopted a canonical radius $r_{2500}$, within which the
mean mass density is 2500 times the critical density of the Universe
at the redshift of the cluster.  (The $r_{2500}$ values are determined
directly from the Chandra data,  with confidence limits calculated from the
$\chi^2$ grids.)  The $r_{2500}$ values are well-matched to the
outermost radii at which  reliable temperature measurements can be
made from the Chandra data.  Note that the data for PKS0745-191 do not
quite reach to $r_{2500}$ and for this cluster we quote $f_{\rm gas}$
at the outermost radius at which reliable  measurements can be made
$\sim 0.8r_{2500}$.  Although independent  confirmation of the  X-ray
mass results for 3C295 is not available, we  include this cluster in
our simple since in most other ways it appears similar  to the other
objects in the sample. The $r_{2500}$ values for the clusters are listed 
in Table \ref{table:fgas}. The best-fit NFW model parameters and 68 per cent 
confidence limits are summarized by Allen \etal (2001b).

\begin{figure}
\vspace{0.5cm}
\hbox{
\hspace{-0.5cm}\psfig{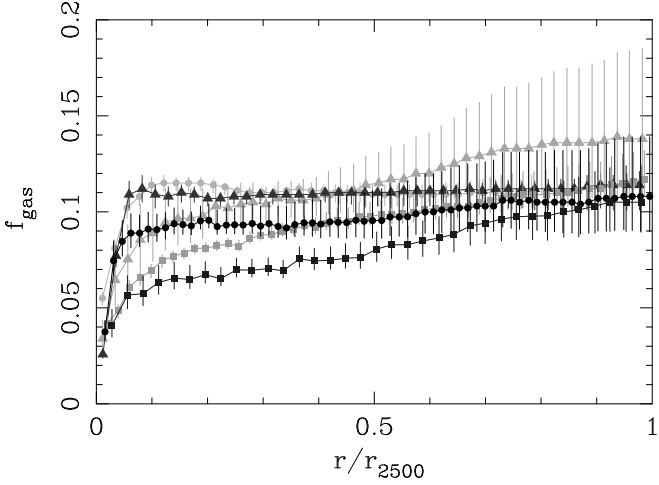}
} \caption{The observed X-ray gas mass fraction profiles  with the
radial axis scaled in units of $r_{2500}$.  Symbols are as follows:
PKS0745-191 (light circles), Abell 2390 (light triangles), Abell 1835
(dark triangles), MS2137-2353 (light squares), RXJ1347-1145 (dark
circles), 3C295 (dark squares). The default $\Lambda$CDM cosmology is
assumed.  Note that $f_{\rm gas}(r)$ is an integrated quantity and so
the  error bars on neighbouring points in a profile are correlated.
}\label{fig:fgas}
\end{figure}

\section{Results}

\begin{table*}
\begin{center}
\caption{The observed X-ray gas mass fractions (and 68 per cent
confidence limits) measured at $r_{2500}$ (in Mpc)
for the default SCDM and $\Lambda$CDM cosmologies. 
%The last two lines 
%list the weighted-mean X-ray gas mass fraction and $\Omega_{\rm m}$ values. 
}\label{table:fgas}
\vskip -0.1truein
\begin{tabular}{ c c c c c c c }
&&&&&&  \\
\multicolumn{1}{c}{} &
\multicolumn{1}{c}{} &
\multicolumn{2}{c}{SCDM} &
\multicolumn{1}{c}{} &
\multicolumn{2}{c}{$\Lambda$CDM} \\
\multicolumn{1}{c}{} &
\multicolumn{1}{c}{} &
\multicolumn{1}{c}{{$r_{2500}$}} &
\multicolumn{1}{c}{{$f_{\rm gas}$}} &
\multicolumn{1}{c}{} &
\multicolumn{1}{c}{{$r_{2500}$}} &
\multicolumn{1}{c}{{$f_{\rm gas}$}} \\
\hline					                                    
PKS0745-191   & ~ & $0.85^{+0.04}_{-0.05}$  & $0.174^{+0.013}_{-0.012}$  &  & $0.68^{+0.03}_{-0.03}$ &  $0.112^{+0.008}_{-0.009}$ \\
Abell 2390    & ~ & $0.69^{+0.14}_{-0.09}$  & $0.209^{+0.060}_{-0.046}$  &  & $0.64^{+0.15}_{-0.09}$ &  $0.138^{+0.047}_{-0.033}$ \\
Abell 1835    & ~ & $0.72^{+0.05}_{-0.03}$  & $0.164^{+0.016}_{-0.016}$  &  & $0.66^{+0.06}_{-0.02}$ &  $0.114^{+0.006}_{-0.013}$ \\
MS2137-2353   & ~ & $0.49^{+0.03}_{-0.01}$  & $0.159^{+0.009}_{-0.016}$  &  & $0.46^{+0.02}_{-0.03}$ &  $0.117^{+0.015}_{-0.009}$ \\
RXJ1347-1145  & ~ & $0.72^{+0.10}_{-0.08}$  & $0.142^{+0.034}_{-0.027}$  &  & $0.73^{+0.08}_{-0.09}$ &  $0.108^{+0.031}_{-0.018}$ \\
3C295         & ~ & $0.42^{+0.03}_{-0.03}$  & $0.128^{+0.020}_{-0.016}$  &  & $0.41^{+0.04}_{-0.03}$ &  $0.105^{+0.019}_{-0.016}$ \\
&&&&&& \\					                                    
%$\bar{f_{\rm gas}}$ & ~ & & $0.1600 \pm 0.0070$                      &  &  &  $0.1131 \pm 0.0053$     \\
%$\Omega_b$         & ~ & & $0.0820 \pm 0.0072$                      &  &  &  $0.0418 \pm 0.0037$     \\
%$\Omega_m$          & ~ & & $0.452  \pm 0.044$                       &  &  &  $0.319  \pm 0.032$      \\
%&&&&&& \\								      								     
\hline									      								     
\end{tabular}								      								     
\end{center}
% \parbox {7in}
%{}
%\vskip 0.5truein
\end{table*}

\subsection{The X-ray gas mass fraction measurements}

Fig. \ref{fig:fgas} shows the observed $f_{\rm gas}(r)$ profiles for
the six clusters assuming the standard $\Lambda$CDM cosmology. We see
that  although some variation is present from cluster to  cluster, the
profiles tend towards a similar value at $r_{2500}$.  Table
\ref{table:fgas} lists the results on the X-ray gas mass fractions
measured at $r_{2500}$ for both the SCDM and $\Lambda$CDM
cosmologies. Taking the weighted mean of the $f_{\rm gas}$ results
for all six clusters studied, we obtain ${\bar f_{\rm gas}} = 0.160 \pm
0.007$ for SCDM (h=0.5) and ${\bar f_{\rm gas}} = 0.113 \pm 0.005$ for 
$\Lambda$CDM (h=0.7).

In calculating the total baryonic mass in the clusters, we assume
that the optically luminous baryonic mass in galaxies is $0.19h^{0.5}$
times the  X-ray gas mass (White \etal 1993; Fukugita, Hogan \&
Peebles 1998).  Other sources of baryonic matter are
expected to make only very small contributions to the total mass 
and are ignored. 

Given the baryonic masses, and assuming that the regions of the clusters 
within $r_{2500}$ provide a fair
sample of the matter content of the Universe, we can write

\begin{equation}
\Omega_{\rm m} =  \frac{\Omega_{\rm b}}{f_{\rm gas}(1+0.19h^{0.5})}.
\end{equation}

\noindent For $\Omega_{\rm b}h^{2} =  0.0205\pm0.0018$ (O'Meara \etal 2001) 
and using the $\Lambda$CDM ($h=0.7$) $f_{\rm gas}$ values, we obtain the 
(self-consistent) result $\Omega_{\rm m} = 0.319 \pm 0.032$. Using 
the SCDM ($h=0.5$) $f_{\rm gas}$ values, we obtain $\Omega_{\rm m} = 
0.452 \pm 0.044$.

\subsection{Cosmological constraints from the $f_{\rm gas}(z)$ data}

\begin{figure*}
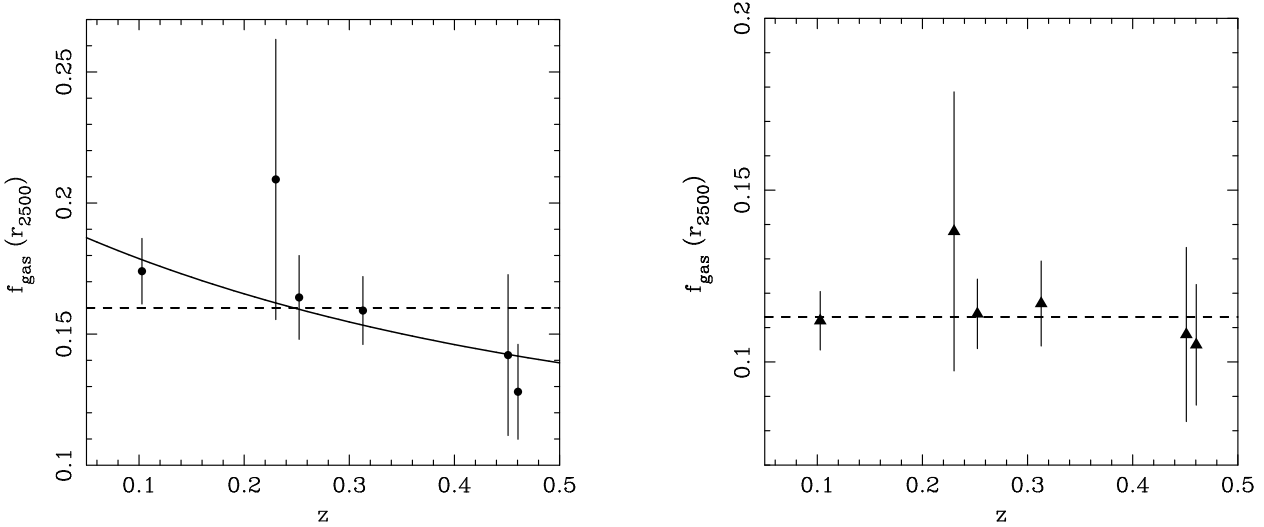

\vspace{0.2cm}
\hbox{
\hspace{0.2cm}\psfig{figure=fig2a.ps,width=0.43 \textwidth,angle=270}
\hspace{1.3cm}\psfig{figure=fig2b.ps,width=0.43 \textwidth,angle=270}
}
\caption{The apparent variation of the observed X-ray gas mass
fraction (with  root-mean-square $1\sigma$ errors) as a  function of
redshift for the default (a: left panel) SCDM (h=0.5) and  (b: right
panel) $\Lambda$CDM (h=0.7) cosmologies.  The dashed curves show the
results of fitting a constant value to the  data in each case. The
solid line in (a) shows the predicted  curve for the best-fit
cosmology with $\Omega_{\rm m}=0.30$ and $\Omega_{\Lambda}=0.95$ (see
Section 3.2).  }\label{fig:redshift}
\end{figure*}

In addition to the simple calculation of $\Omega_{\rm m}$ based on the
weighted-mean $f_{\rm gas}$ values, described above, the data for the
present sample can be used to obtain more rigorous constraints on
cosmological parameters from the apparent variation of $f_{\rm gas}$
with redshift.

Fig. \ref{fig:redshift}~shows the $f_{\rm gas}$ values   as a
function of redshift for the SCDM and $\Lambda$CDM cosmologies.  We
see that whereas the results for the $\Lambda$CDM  cosmology are
consistent with a constant $f_{\rm gas}$ value, the results  for SCDM
indicate an apparent drop in $f_{\rm gas}$ as the redshift increases.
The differences in the $f_{\rm gas}(z)$ behaviour for the SCDM and
$\Lambda$CDM cosmologies reflect the dependence of the $f_{\rm
gas}(z)$  measurements on the assumed angular diameter distances to
the clusters  ($f_{\rm gas} \propto D_{\rm A}^{1.5}$). Under the
assumption that the  $f_{\rm gas}$ values should be invariant with
redshift, as would be expected  if rich, relaxed clusters provide a
fair sample of the matter content of the  Universe, we can see from
inspection of Fig. \ref{fig:redshift} that the data for the 
present sample favour the $\Lambda$CDM over the SCDM cosmology.

In order to quantify more precisely the degree to which our data  can
constrain the relevant cosmological parameters, we have  fitted the
data in Fig. \ref{fig:redshift}(a) with a model which accounts  for 
the expected apparent variation in the $f_{\rm gas}(z)$ values, which are 
measured assuming an SCDM cosmology, for different underlying 
cosmologies. The `true' cosmology should be the
cosmology that provides the best fit to the measurements. (We work
with the SCDM data. Note that the $f_{\rm gas}(r)$ profiles exhibit 
only small variations around $r_{2500}$, and so the effects of changes in 
$r_{2500}$ as the cosmology is varied can be ignored.) 

The model function fitted to the data is 

\begin{eqnarray}
f_{\rm gas}^{\rm mod}(z) =
\frac{
\Omega_{\rm b}} {\left(1+0.19
\sqrt{h}\right) \Omega_{\rm m}} \left[ \frac{h}{0.5} \, \frac{D_{\rm A}^{\Omega_{\rm m}=1, \Omega_{\rm 
\Lambda}=0 }(z)}{D_{\rm A}^{\Omega_{\rm m},\,\Omega_{\Lambda}}(z)}
\right]^{1.5}, 
\end{eqnarray}

\noindent which depends on $\Omega_{\rm m}$, $\Omega_{\Lambda}$,  $\Omega_{\rm
b}$ and $h$.  The ratio $(h/0.5)^{1.5}$ accounts for the change in the
Hubble Constant between the considered model and default SCDM 
cosmology, and the ratio of the angular diameter distances accounts
for deviations in the geometry of the Universe from the Einstein-de 
Sitter case. We constrain $\Omega_{\rm b}h^{2} =  0.0205\pm0.0018$
(O'Meara \etal 2001) and $h=0.72\pm0.08$, the final result  from the
Hubble Key Project reported by Freedman \etal (2001).  The $\chi^2$
difference between the model and SCDM data is then

\begin{eqnarray}
\chi^2& =& \sum_{\rm all\;clusters}
\frac{\left[f_{\rm gas}^{\rm mod}(z_{\rm i})- f_{\rm gas,\,i}
\right]^2}{\sigma_{f_{\rm gas,\,i}}^2}\nonumber\\
&&+\left(\frac{\Omega_{\rm
b}h^2-0.0205}{0.0018} \right)^2 +\left(\frac{h-0.72} {0.08} \right)^2,
\end{eqnarray}

\noindent where $f_{\rm gas,\,i}$ and $\sigma_{f_{\rm gas,\,i}}$ are
the  best-fit values and symmetric root-mean-square errors for the
SCDM data from Table \ref{table:fgas}, and $z_{\rm i}$ are the redshifts 
of the clusters. We have examined a grid of cosmologies
covering the plane $0.0< \Omega_{\rm m}  < 1.0$ and $0.0< \Omega_{\rm
\Lambda} < 1.5$. The joint  1, 2 and 3 $\sigma$ confidence contours on
$\Omega_{\rm m}$ and $\Omega_{\rm \Lambda}$ (corresponding to $\Delta
\chi^2$ values of 2.30, 6.17  and 11.8, respectively) determined from
the fits are shown in Fig. \ref{fig:cosmo}.

The best-fit cosmological parameters and marginalized 1$\sigma$ error
bars  are $\Omega_{\rm m} = 0.30^{+0.04}_{-0.03}$  and
$\Omega_{\Lambda} = 0.95^{+0.48}_{-0.72}$, with  $\chi^2_{\rm
min}=1.7$ for 4 degrees of freedom, indicating that the model provides
an acceptable description of the data.  The best-fit cosmological
parameters are similar to  those assumed for the default $\Lambda$CDM
cosmology in  Fig. \ref{fig:redshift}b, which is expected given the 
approximately constant nature of the $f_{\rm gas}$(z) values shown in 
that Figure.

\section{Discussion}

\begin{figure*}
\vspace{0.5cm}
\hbox{
\hspace{0.2cm}\psfig{figure=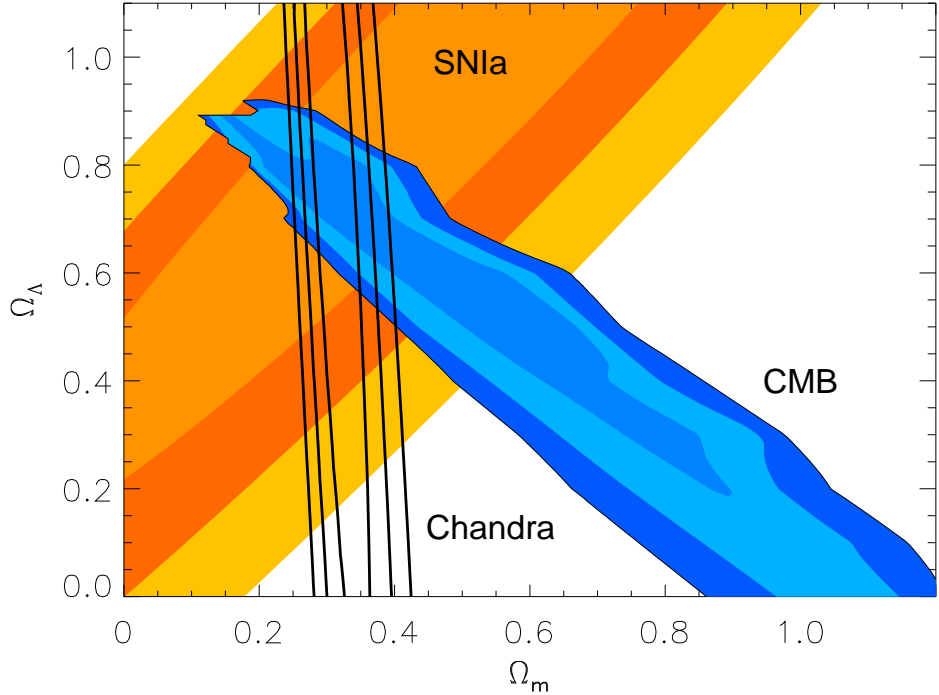,width=.85\textwidth,angle=270}
}
 \caption{The joint 1, 2 and 3 $\sigma$ confidence contours on 
$\Omega_{\rm m}$ and $\Omega_{\Lambda}$ determined from the 
Chandra $f_{\rm gas}$(z) data (bold contours), and independent analyses 
of cosmic microwave background (CMB) anisotropies 
and the properties of distant supernovae (from Jaffe \etal 2001).}
\label{fig:cosmo}
\end{figure*}

The result on the mean matter density of the Universe, $\Omega_{\rm m}
=0.30^{+0.04}_{-0.03}$, determined from the Chandra results on the
X-ray gas mass fraction for the present sample of relaxed, lensing
clusters, represents one of the tightest constraints  on this
cosmological parameter to date.  The variation of the gas mass
fraction with redshift also yields the measurement of a  positive
cosmological constant with $\Omega_{\rm \Lambda} =
0.95^{+0.48}_{-0.72}$, in good agreement with previous results based
on studies of the properties of distant supernovae (Riess \etal 1998;
Perlmutter \etal 1999)

In Fig. \ref{fig:cosmo} we show a comparison of the joint constraints  on
$\Omega_{\rm m}$ and $\Omega_{\Lambda}$  determined from the Chandra
$f_{\rm gas}(z)$ data, with the results of Jaffe \etal (2001) from
studies of cosmic microwave background (CMB) anisotropies
(incorporating the COBE  Differential Microwave Radiometer,
BOOMERANG-98 and MAXIMA-1 data of Bennett \etal 1996, de Bernardis
\etal 2000 and Hanany \etal 2000, respectively)\footnote{We note that
the  results on $\Omega_{\rm m}$ and $\Omega_{\rm \Lambda}$ from the
CMB data  reported by Jaffe \etal (2001) are consistent with, though
less constraining  than, the later analyses of de Bernardis \etal
(2002) and  Stompor \etal (2001)  using the full BOOMERANG and
MAXIMA-1 data sets.} and the properties of distant supernovae
(incorporating the data of Riess \etal 1998 and Perlmutter \etal
1999).  The agreement between the results obtained from the
independent  methods is striking: all three data sets are consistent,
at the $1\sigma$  confidence level, with a cosmological model with
$\Omega_{\rm m} = 0.3$  and $\Omega_{\Lambda} = 0.7-0.8$. These
results are also consistent  with the findings of Efstathiou \etal
(2001) from a combined analysis  of the 2dF Galaxy Redshift Survey and
CMB data.

An important aspect of the present work is that, in addition to the
exquisite data quality provided by Chandra, the clusters studied are
all regular, relatively relaxed systems for which independent
confirmation  of the mass results is available from
gravitational lensing studies. The systematic uncertainties in the
$f_{\rm gas}$ measurements are therefore greatly reduced with respect
to  previous X-ray studies.  For both Abell 2390 and RXJ1347-1145, the
X-ray and weak lensing mass profiles are consistent within their 68
per cent  confidence limits. For Abell 1835, 2390, MS2137-2353
and PKS0745-191, the  observed strong lensing configurations (on
scales $r \sim 20-80\,h^{-1}$kpc)  can be explained by mass models
within the 68 per cent Chandra confidence  contours, although redshift
measurements for the arcs (which are required  to define the lensing
masses precisely) are not available in all  cases.\footnote{For
RXJ1347-1145, a two-component mass  model, consistent  with the
complex X-ray structure observed in the   southeast quadrant, is
required to explain the strong lensing data.}  The presence of
significant  non-thermal pressure support (\eg arising from turbulent
and/or bulk  motions and/or magnetic fields) on scales $\sim r_{2500}$
can therefore be excluded, and the residual systematic uncertainties in the
$f_{\rm gas}$ values are small ($\approxlt 10$ per cent {\it i.e.}
smaller, typically, than the statistical uncertainties. We note 
that the effects of departures from spherical symmetry on the 
$f_{\rm gas}$ results are expected to be $\approxlt$ a few per cent \eg
Buote \& Canizares 1996).

The observed $f_{\rm gas}(r)$ profiles are essentially flat  around
$r_{2500}$, which supports the assumption that the measured $f_{\rm
gas}$ values represent a fair sample of the matter content of the
Universe.  If, however, the values were to rise by a further $\sim 10$
per cent beyond $r_{2500}$, the result on $\Omega_{\rm m}$ would drop
by a corresponding amount.  The $f_{\rm gas}$ values measured at
$r_{2500}$ are not sensitive to the choice of using an NFW model to
parameterize  the total mass distributions in the clusters. Repeating
the  analysis presented here using either a non-singular isothermal
sphere or a Moore \etal (1998) model to parameterize the total mass
distributions leads to  results on the weighted-mean $f_{\rm gas}$
values in good agreement with those quoted in Section 3.1. We note,
however, that the $f_{\rm gas}(r)$ profiles determined using the
different mass models exhibit some systematic variation, particularly
at small radii  ($r\approxlt 0.1r_{2500}$), and when extrapolated to
large radii ($r > r_{2500}$), as can be expected given the different
asymptotic slopes.  In a future paper, we will examine in detail the
ability of different  parameterized models to describe the Chandra
data for relaxed clusters.

The constraints on $\Omega_{\rm m}$ and $\Omega_{\Lambda}$  should
improve as further Chandra, XMM-Newton and high-quality  gravitational
lensing data become available for more regular,  relaxed clusters,
especially at high redshifts (although relaxed  systems, like those
studied here, are expected to be  very rare at high redshifts).  This
work can also be extended to include less relaxed clusters, or
clusters  which appear relaxed at X-ray wavelengths but for which
independent confirmation  of the mass results from gravitational
lensing studies is not yet  available, although this will require
careful consideration of the  additional systematic uncertainties
involved.

We thank Andrew Jaffe for providing the CMB and supernovae results shown 
in Fig.~\ref{fig:cosmo}. SWA and ACF thank the Royal Society for support.

\end{document}